\documentclass[doublecol]{epl2} 
\usepackage{epstopdf}

\title{A new insight on stellar age I. Theoretical trends with gyro-age and rotation}
\shorttitle{A new insight on stellar age} %Insert here a short version of the title if it exceeds 70 characters

\author{D. B. de Freitas}
\shortauthor{D. B. de Freitas}

\institute{Departamento de F\'{\i}sica, Universidade Federal do Cear\'a, Caixa Postal 6030, Campus do Pici, 60455-900 Fortaleza, Cear\'a, Brasil\\
INAF-Osservatorio Astrofisico di Catania, Via S. Sofia 78, I-95123, Catania, Italy}

\abstract{In this first work attempts to analytically explain the effects on the magnetic braking index, $q$, caused by the evolution of stellar velocity in main-sequence stars, and estimated by de Freitas \& De Medeiros (2013). We have found that the effect of $q$ is here a determining factor for understanding the delicate mechanisms that control the spin-down of stars as a function of the mass of stars. We note that our models predict that the calculated ages are distinct from gyrochronology ages. Indeed, the gyro-ages are measured considering only the canonical value of the Skumanich relation ($q$=3). As a result, we find that the age of stars can be well-determined when $q$ is free parameter. We also verified that for rotation periods less than $\sim$ 5 days (i.e., fast rotators) there is a strong discrepancy among the different indexes $q$. In addition, the ages measured by gyrochronology model can be underestimated according to mass range selected. In conclusion, we suggest that the generalized gyro-ages by magnetic braking index can be an interesting way to better understand the idea of rotation as a clock.}

\begin{document}

\maketitle

\section{Introduction}

It is widely accepted that magnetic braking is a fundamental concept for understanding angular momentum losses due to magnetic stellar winds for several classes of stars, such as main-sequence field and cluster stars. This mechanism was initially suggested by Schatzman \cite{Schatzman}, who pointed out that slow rotators have convective envelopes. As mentioned by Kraft \cite{Kraft}, the behaviour of the mean rotational velocity of low-mass-main-sequence stars below 1.5$M_{\odot}$ (spectral type F0) is preferentially due to magnetic wind. A few years later, Skumanich\cite{sku}'s pioneering work argued that stellar rotation, activity, and lithium abundances for solar-like stars obeys= a simple relationship given by $\mathrm dJ/\mathrm dt \propto \Omega^{3}$, where $t$ is time, $J$ is the angular momentum and $\Omega$ denotes the angular velocity. These authors established early on that stellar rotation and age should be related in cool main-sequence stars. The next step was to develop a more complete theory capable of explaining how losses of angular momentum occurs.

Inspired by \cite{mestel1968,mestel1984,mestel1987}'works, Kawaler \cite{kawaler1988} elaborated on a theoretical model for describing the behaviour of the loss of angular momentum for main-sequence stars with masses less than 1.5$M_{\odot}$ due to wind ejected by stars. This wind gets caught by the magnetic field that spins outward until it is ejected, affecting the angular momentum and causing slowdown. In this way, the magnetic field acts like a brake.

As the magnetic field strength depends on stellar mass, Chaboyer \textit{et al.}\cite{chaboyer1995} modified the Kawaler's parametrization and introduced a saturation level into the angular momentum loss law. On the other hand, \cite{kris1997} proposed the inclusion of a Rossby scaling at the saturation velocity for stars more massive than 0.5$M_{\odot}$. More recently, the Roosby number (defined as the ratio between the rotational period and the convective overturn timescale) is defined to characterize the deviation from the Skumanich law.

In this context, the rotation--age relationship offers a promising method for measuring the ages of field stars, providing the rotation as an attractive alternative chronometer \cite{epstein2012}. Of the methods and calibrations to estimate the ages of stars, the present work will focus on the gyrochronology.

Barnes \cite{barnes2003} proposed a simple formulation that uses the colour and period values to derive the stellar ages for solar- and late-type stars. According to authors, the age and colour dependences of sequences of stars allow us to identify their underlying mechanism, which appears to be primarily magnetic. This determination of stellar ages from their rotational periods and colours, he named ``Stellar Gyrochronology''. Gyrochronology is a age-dating technique that provides precise and accurate ages for low mass (G, K and M) stars on the main-sequence \cite{gallet}. The theoretical background is centred on a functional formulae based on the Skumanich-type age dependence (square root of time) fitted by the following expression\begin{equation}
\label{int1}
P=g(t)f(B-V),
\end{equation} 
where $P$, $t$, and $B-V$ are the rotational periods (days), ages (Myr), and colours, respectively. The function $g(t)\propto t^{1/2}$ is the rotation-age relation from \cite{sku}, However, only a sequence of stars, also called sequence $I$, respects this condition. This sequence consists of stars that form a diagonal band of increasing periods with increasing $B-V$ colours in a colour period diagram. There is also another sequence of stars denoted by the letter $C$ that represents the fast rotators. For these stars, the magnetic field is expected to be saturated. Consequently, there is no clear dependence between the period and colour. In this case, the expression for the angular momentum loss rate is given by $\mathrm dJ/\mathrm dt \propto \Omega$, and therefore, the period and age are related by a simple exponential law \cite{chaboyer1995,barnes2003,barnes2007,pacepas,defreitasetal15}.

Recently, de Freitas \& De Medeiros \cite{defreitas2013} revisited the modified Kawaler parametrization proposed by Chaboyer \textit{et al.} \cite{chaboyer1995} in the light of the nonextensive statistical mechanics \cite{tsallis1988}. de Freitas \& De Medeiros \cite{defreitas2013} analyzed the rotational evolution of the unsaturated F- and G- field stars that are limited in age and mass within the solar neighbourhood using a catalogue of $\sim$16000 stars in the main sequence \cite{holmberg2007}. More recently, de Freitas et al. \cite{defreitasetal15} generalized the Reiners \& Mothanty \cite{reiners2012}'s torque using the nonextensive framework. In both cases, they use the $q$-index extracted from the Tsallis formalism as a parameter that describes the level of magnetic braking. They also linked this parameter with the exponent of dynamo theory ($a$) and with magnetic field topology ($N$) through the relationship $q=1+\frac{4aN}{3}$, where $q$ is defined as the stellar magnetic braking index. As a result, they showed that the saturated regime can be recovered in the nonextensive context, assuming the limit $q\rightarrow1$. This limit is particularly important because it represents the thermodynamic equilibrium valid in the Boltzmannian regime. According this work, the torque in the generalized version is given as $\mathrm dJ/\mathrm dt \propto \Omega^{q}$, indicating that the rotational velocities of F- and G-type main-sequence stars decrease with age according to $t^{1/(1-q)}$. The values of $q$ obtained by de Freitas \& De Medeiros \cite{defreitas2013} suggest that it has a strong dependence on stellar mass.

In a recent paper, van Saders  {\it et al.} \cite{van} discuss the validity of the gyrochronological model for stars older than the Sun. They show that for ages greater than that of our Sun, a fundamental change occurs in the magnetic nature of the stars, revealing a critical transition that separates the strength of the stellar winds into two regimes. In short, the magnetic braking weakens as the star's age advances. With this knowledge, the authors point out that the use of gyrochronology for older stars in the main sequence becomes limited when weakened magnetic braking begins. To this end, they show that this behaviour occurs for a selected sample of 21 stars observed by Kepler mission. In particular, this result reveals another important point for the present study: we cannot consider the same magnetic braking law for the lifetimes of the stars in the main sequence or even that stars must have the same magnetic braking mechanism as the Sun.  

The main goal of the research is to show that rotation does not account for itself when explaining the star clock. It is necessary to verify a fine structure that can be responsible via the variation of the magnetic braking index marked by changes of the moment of inertia. Undoubtedly, the key feature here is the variation of the braking index, which cannot be held constant but instead evolves and diminishes on timescales shorter that the stellar life span. Nevertheless, this assumption will be discussed in the second paper in this series. In this first work, we present a new insight for understanding the behaviour of the magnetic braking index and its correlation with the gyrochronology ages (hereafter gyro-ages) developed by Barnes\cite{barnes2010a}.  

\begin{figure*}
	\begin{center}
		\includegraphics[width=0.99\textwidth]{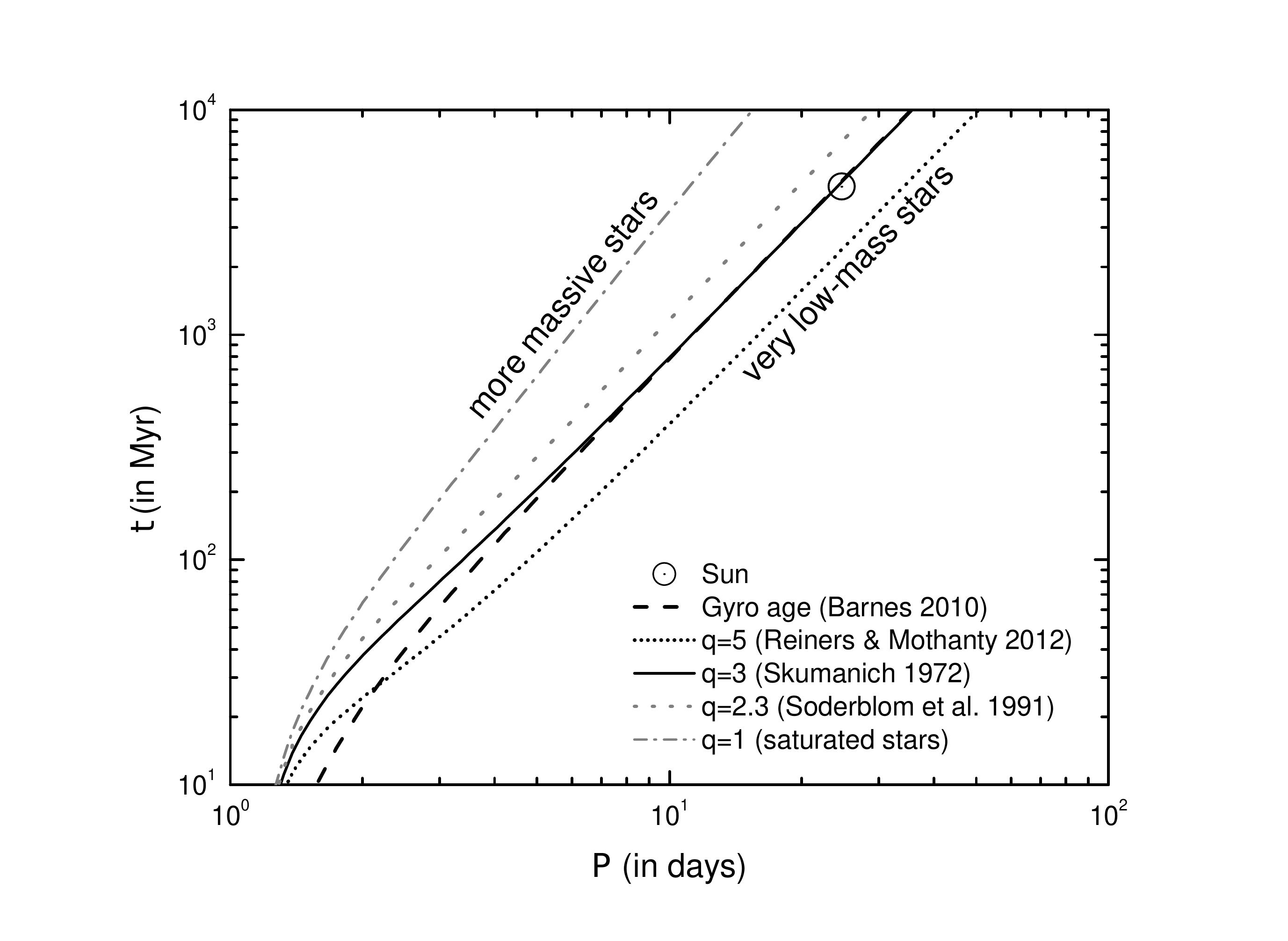}
	\end{center}
	\caption{Stellar age $t$ (in Myr) as a function of period $P$ (in days) for differing initial conditions of braking index $q$ based on the stellar mass. According to de Freitas \& De Medeiros \cite{defreitas2013}, the entropic index $q$ is a function of mass and, therefore, when $q$ increases the mass descreases. In particular, $q$-index for stars in the saturated regime corresponding to 1, $q=2.3$ denotes the Soderblom \textit{et al.} \cite{Soderblom1991}'s exponent, $q=$3 is related to Skumanich index and $q=$5 for the very-low mass stars as propused by Reiners \& Mohanthy \cite{reiners2012}. In our model, we used $P_{0}$=1.1 days for solar calibration as indicated by Barnes \cite{barnes2010}. The Sun is indicated by symbol $\odot$ with rotation period of 25 days. Figure also shows that the $q$-index is a function of mass as proposed by de Freitas \& De Medeiros \cite{defreitas2013}.}
	\label{fig1}
\end{figure*}

\section{Generalized gyro-ages}
By using a nonextensive framework \cite{tsallis1998,tsallis1999,tsallis2004,defreitas2012,Silvaetal13}, de Freitas \& De Medeiros \cite{defreitas2013} proposed the following equation to describe the behaviour of the rotational velocity of the main-sequence stars:
\begin{equation}
\begin{array}{cc}
\dot{\Omega}=-\lambda_{q}\Omega^{^{q}}, \quad q\geq 1
\end{array}
\label{4}
\end{equation}
where $\lambda_{q}$ denotes the braking strength. In the literature, the $q$-index is limited to the range of 1 (stars with saturated magnetic field and, therefore, more massive ones) to 5 (corresponding to the Reiners \& Mothanty relation for the very low-mass stars).

Stellar age is the more important stellar parameter for investigating the spin-downs of stars. We have shown in eq. (\ref{4}) that the spin-down rate is proportional to a power law that decays as a function of the magnetic braking index $q$. In general, the solution of the referred equation for the ages of stars, $t$, is given by
\begin{equation}
\label{sa1}
t_{q}=\frac{1}{q-1}\frac{\Omega}{\dot{\Omega}}\left[1-\left(\frac{\Omega}{\Omega_{0}} \right)^{q-1}  \right] ,\quad q>1
\end{equation}
and
\begin{equation}
\label{sa1x}
t_{1}=\frac{\Omega}{\dot{\Omega}}\ln\left(\frac{\Omega}{\Omega_{0}}\right)  ,\quad q=1
\end{equation}
where $\Omega_{0}$ is the angular velocity at time $t=0$ and $\Omega$ is the velocity at time $t=t_{age}$, i.e., now. As quoted by de Freitas \& De Medeiros \cite{defreitas2013}, $q>1$ denotes the unsaturated magnetic regime, whereas $q=1$ is the index associated to the saturated regime. 

A later equation would give the star's ``true'' age if only the magnetic torque is responsible for the star spin-down and the $q$-index is a constant. Basically, this generalization is a similar age-dependent factor $g(t)$ in eq. (\ref{int1}), described by a power law of $t^{n}$, where $n=1/(q-1)$.

The solution was only possible because we have assumed that $\lambda_{q}$ and the $q$-index remain constant with time, i.e., the star's moment of inertia does not change over time. In addition, it is easy to show that the stellar age (see eq. \ref{sa1}) increases monotonically with decreasing $q$ (including the special case of $q=1$). As quoted by de Freitas \& De Medeiros \cite{defreitas2013}, according to the stellar mass, larger values of $q$ mean that the star spins down more slowly.

A good test is to deduce the gyro-ages extracted from Barnes \cite{barnes2010a}. These ages can be recovered using the above expression (\ref{sa1}) and eq. (2) from Barnes \cite{barnes2010a}'s paper, given by
\begin{equation}
\label{sa2a}
\frac{P}{\dot{P}}=\left(\frac{k_{I}P^{2}}{\tau}+\frac{\tau}{k_{C}}\right),
\end{equation}
where $\tau$ is the convective turnover timescale, and $k_{C}$ and $k_{I}$ are two dimensionless constants whose values can be derived from observations, approximately $k_{C}$=0.646 days Myr$^{-1}$ and $k_{I}$=452 Myr day$^{-1}$. However these values must be satisfied considering the e-folding timescale of 56 Myr for a 1M$_{\odot}$ star and the rotation period started off with an initial period of 1.1 days for a star with solar age and period. Parameters $P$ and $\tau$ are formulated in terms of the Rossby number, $R_{0}$. Barnes \cite{barnes2010a} used a more long way to estimate a relation between age and rotation period. Our goal is to find an alternative and easier way using only eq. (\ref{4}) used by \cite{defreitas2013}.

To this end, we use the Skumanich index, $q=3$ and $\frac{\Omega}{\dot{\Omega}}=-\frac{P}{\dot{P}}$ obtained using the relation $\Omega=\frac{2\pi}{P}$. Thus, we found that
\begin{equation}
\label{sa2}
t_{q=3}=\left(\frac{\tau}{k_{C}}\right)\left[\frac{1}{2}\left(1-\frac{P^{2}_{0}}{P^{2}}\right)\right]+\left(\frac{k_{I}}{2\tau}\right)\left(P^{2}-P^{2}_{0}\right).
\end{equation}
where this equation is similar to equation (32) from \cite{barnes2010a} 
\begin{equation}
\label{barnes}
t=\left(\frac{\tau}{k_{C}}\right)\ln\left(\frac{P}{P_{0}}\right)+\left(\frac{k_{I}}{2\tau}\right)\left(P^{2}-P^{2}_{0}\right).
\end{equation}

Equation (\ref{barnes}) is derived from nonlinear model for rotation of cool stars, i.e., stars with mass less than the solar mass. Nevertheless, it assumes that there is a same braking law for these stars. de Freitas \& De Medeiros \cite{defreitas2013} showed that the $q$-index which controlls magnetic braking is a function of mass and, therefore, $q=3$ is not able to estimate good ages for stars outside their valid range. Thus, $q=3$ is restricted to low-mass stars. According to de Freitas \& De Medeiros \cite{defreitas2013}, stars colder than the Sun must have the $q$-index greater than 3 for solar-type stars with spectral type G and higher $q$ lower than 3 for more massive stars as indicated by Fig. \ref{fig1}. An important point is that the trend obtained from eqs. (\ref{sa2}) and (\ref{barnes}) are useful to describe the behaviour of the disc-less stars. On the other hand, accreting stars has the physics more complex than that of the non-accreting stars, because the star-disc interation controls the rotation \cite{vidotto}.

Figure \ref{fig1} illustrates the behaviour between eqs. (\ref{sa2}) and (\ref{barnes}). This figure show that they are similar for periods greater than $\sim$ and $q=3$. In fact, the equations differ only in the first term on the right side. Compared to the second term (where both equations are equal), the first term has a very small contribution, affecting only short periods. In this way, the profiles of the curves $q=3$ and the gyro-age shown in Fig. \ref{fig1} are dominated exclusively by the second term. 

On the other hand, using equation (\ref{sa1x}), we can obtain the following expression for stars in the saturated magnetic field regime \cite{chaboyer1995} 
\begin{equation}
\label{sa3}
t_{q=1}=\frac{P}{\dot{P}}\ln\left(\frac{P}{P_{0}}\right),
\end{equation}
where $\frac{P}{\dot{P}}$ is given by eq. (\ref{sa2a}). Therefore, we have
\begin{equation}
\label{sa4}
t_{q=1}=\left(\frac{k_{I}P^{2}}{\tau}+\frac{\tau}{k_{C}}\right)\ln\left(\frac{P}{P_{0}}\right).
\end{equation}

In Figure \ref{fig1}, stellar age $t$ (in Myr) is shown as a function of period $P$ (in days) for different values of $q$. Indeed, de Freitas \& De Medeiros \cite{defreitas2013} have shown that the $q$-index is a function of mass and when $q$ increases the mass descreases. For stars in the saturated regime, $q$-index corresponding to 1.  Soderblom et al. \cite{Soderblom1991} used $q$=2.3 for stars of the lower main sequence where the chromospheric emission-age relation was observed. Skumanich relation used for solar type stars is recovered when $q=$3. Finally, $q=$5 for the very-low mass stars as proposed by Reiners \& Mohanthy \cite{reiners2012}. In our model, we used $P_{0}$=1.1 days for solar calibration as indicated by Barnes \& Kim \cite{barnes2010a}. The Sun is indicated by symbol $\odot$ with rotation period of 25 days. Figure \ref{fig1} also shows that the $q$-index is a function of mass as proposed by de Freitas \& De Medeiros \cite{defreitas2013}. 

\section{Concluding remarks}
Undoubtedly, the strongest conclusion in this study concerns the role of the $q$-index as an important parameter to explain the mechanism that controls stellar clock as a function of mass. In this sense, $q$ is a parameter sensitive to the stellar mass that determine how strong magnetized winds are. In addition, this parameter provides us with a powerful diagnostic tool concerning the "rotational lives" of the stars on the main sequence. In addition, the role of the $q$-index on the determination of the stellar ages is an important step to improve our understanding of the regimes close to the unsaturated regime of the magnetic field, where are found the more massive stars.

Given the strong agreement of our model with the previously estimated ages by gyrochronology for stars with rotation period greater than 5 days, we suggest that the magnetic braking index ($q$) can be an interesting and alternative way to better understand the idea of stellar rotation as an astronomical clock. Nevertheless, there is a limitation on the conclusions that we can extract from standard gyrochronology scenario: it does not take into account that the index ($q=3$) used is a function of mass and, therefore, as seen in the Fig.~\ref{fig1}, the ages estimated by this model may both overestimate or underestimate the age of the stars. 

Finally, the present scenario shows that both models (generalized and standard gyro-model) are not considered the variation of the moment of inertia. In the next paper, the $q$-index will be estimated for other evolutionary stages, including, for instance, globular and open clusters as well as giant stars, where the effects of the variations of stellar radii should be considered. Future missions, such as PLATO, will be ideally suited to derive accurate stellar ages and will allow us to improve magnetic braking models and age-rotation relationship.

\acknowledgments
DBdeF acknowledges financial support 
from the Brazilian agency CNPq-PQ2 (rant No. 311578/2018-7). Research activities of STELLAR TEAM of Federal University of Cear\'a are supported by continuous grants from the Brazilian agency CNPq.

\end{document}